\documentclass{PoS}

\usepackage[figuresright]{rotating}
\usepackage{graphicx}  % needed for figures
\usepackage{dcolumn}   % needed for some tables
\usepackage{bm}        % for math
\usepackage{amssymb}   % for math
\usepackage{amsmath}
\usepackage{slashed}
\usepackage{stackrel}
\usepackage{feynarts}
\usepackage{color}

\allowdisplaybreaks

\newcommand{\beq}{\begin{equation}}
\newcommand{\eeq}{\end{equation}}
\newcommand{\bqa}{\begin{eqnarray}}
\newcommand{\eqa}{\end{eqnarray}}

\newcommand{\nn}{\pagebreak[0] \nonumber \\}

\def\sam{{{\sc S@M}}}

\def\C++{{{\sc c++ }}}
\def\Feynarts{{{\sc FeynArts}}}
\def\Feyncalc{{{\sc FeynCalc}}}

\newcommand{\QQ}{Q}
\newcommand{\GG}{G}

\newcommand{\GA}{\Gamma}

\newcommand{\padova}{Dipartimento di Fisica e Astronomia, Universit\`a di Padova, and INFN  \\   Sezione di Padova, via Marzolo 8, 35131 Padova, Italy}

\title{Generalised unitarity for dimensionally regulated amplitudes within FDF}

\ShortTitle{Generalised unitarity for dimensionally regulated amplitudes within FDF}

\author{\speaker{William J. Torres Bobadilla}\thanks{Talk based on a collaboration with P. Mastrolia, A. Primo and U. Schubert.}\\
\padova\\
        E-mail: \email{william.torres@pd.infn.it}}

\abstract{
We review the Four-Dimensional-Formulation variant of the Four-Dimensional-Helicity scheme, by showing two applications of this regularisation scheme. The first one is the computation of one-loop helicity amplitudes, for which we present preliminary results for the analytic expressions of the one-loop Higgs plus five- gluon amplitudes. In the second part, we study the Colour-Kinematics duality for off-shell diagrams in gauge theories coupled to matter, showing in a diagrammatic way that the Jacobi relations for the kinematic numerators of off-shell diagrams, built with Feynman rules in axial gauge, reduce to definite set of violating terms due to the contributions of sub-graphs only.
}

\FullConference{12th International Symposium on Radiative Corrections (Radcor 2015) and LoopFest XIV (Radiative Corrections for the LHC and Future Colliders)\\
		15-19 June, 2015\\
		UCLA Department of Physics \& Astronomy Los Angeles, USA}

\begin{document}

\section{Introduction}
\label{sec:intro}
With the successful results delivered by Run I, it is necessary to achieve more accurate results for measure quantities at the theoretical level. Then, in order to test phenomenological predictions we make use of scattering amplitudes, which can be studied in terms of their symmetries and analytic properties. Tree-level or Leading-Order (LO) computations provide a qualitative information, affected by large uncertainties due to the poor convergence of the coupling constant. Therefore, to establish a proper comparison between theory and data, Next-to-Leading-Order (NLO) is needed.\\
As a main ingredient of the NLO contributions we consider one-loop corrections, in which any amplitude can be decomposed in a explicit set of Master Integrals (MI's),  where the coefficients appearing in this combination are  rational functions of the kinematic variables~\cite{Passarino:1978jh}. It is possible to recover the structure of scattering amplitudes
at integral level by constructing the integrands through the multi-particle
pole expansion rising from the analiticity properties and unitarity
of the S-matrix. In fact, scattering amplitudes analytically continued to complex
momenta, reveal their singularity structures in terms of poles and branch
cuts. The unitarity based method (UBM) allows to determine the coefficients of the MI's by
expanding the integrand of tree level cut amplitudes into an expression
that resembles the cut of the basis integrals. 

%%%%%%%%%%
In this talk, we review the four dimensional formulation (FDF) proposed in~\cite{Fazio:2014xea}, in which the Four-Dimensional-Helicity (FDH) scheme~\cite{Bern:1991aq,Bern:1995db,Bern:2002zk} is extended by considering ingredients in four dimensions and providing explicit representations of the polarisation and helicity states for the four-dimensional particles propagating in the loop. FDF has been successfully applied to reproduce one-loop corrections to
$gg \to gg$, $q {\bar q} \to gg$, $gg \to Hg$ (in the heavy top
limit), as well as $gg \to ggg$ and $gg \to gggg$ \cite{Bobadilla:2015wma}.

In addition, we study the Colour-Kinematics (C/K) dual representation of QCD amplitudes by considering the kinematic part of the numerators, which is found to obey Jacobi identities 
and anti-symmetry relations similar to the ones holding for the
corresponding structure constants of the Lie algebra~\cite{Bern:2008qj,Bern:2010ue}.

\par\noindent We report the results obtained in~\cite{Mastrolia:2015maa}, where we consider the tree-level diagrams for $gg\to X$, for massless final state particles, with $X = ss, q\bar{q}, gg$, in four- and $d$- dimensions. We work in axial gauge, describing scalars in the adjoint representation and fermions in the fundamental one. We deal with the Jacobi relation of the kinematic numerators keeping the partons off-shell. Due to the off-shellness of the external particles, the C/K-duality is broken, and anomalous terms emerge. This anomaly vanishes in the on-shell limit, as it should, recovering the exact C/K-duality\\
%%%%%%%%%%%
\section{Four-Dimensional-Formulation}
\label{AppA}
In this section we briefly recall the main features of the FDF scheme.
\begin{itemize}
	\item We use barred notation for quantities referred to unobserved particles,  living in a $d$-dimensional space. Thus, the metric tensor 
	\begin{align}
		\bar g^{\mu \nu} = g^{\mu \nu} + \tilde g^{\mu \nu}  \, ,
	\end{align}
	can be decomposed in terms of a four-dimensional tensor  $g$ and a $-2 \epsilon$-dimensional one, $ \tilde g$.
	The  tensors $g$ and $\tilde g$  project a $d$-dimensional vector $\bar q$ into the four-dimensional  and the  
	$-2 \epsilon$-dimensional subspaces, respectively. 
	\item $d$-dimensional  momenta $\bar \ell$ are decomposed as
	\begin{align}
	\bar \ell = \ell + \tilde \ell \, , \qquad \bar \ell^2  = \ell^2 -\mu^2 = m^2  \,  
	\label{Eq:Dec0}
	\end{align}
	\item The algebra of matrices $\tilde \gamma^\mu = \tilde g^{\mu}_{\phantom{\mu} \nu} \, \bar \gamma^\nu$,
		\begin{align}
			[ \tilde \gamma^{\alpha}, \gamma^{5}   ] &= 0 \, , & 
			\{\tilde \gamma^{\alpha}, \gamma^{\mu}  \} &=0 \ , \label{Eq:Gamma01} &
			\{\tilde \gamma^{\alpha}, \tilde \gamma^{\beta}  \} &= 2 \,  \tilde g^{\alpha \beta} \, . 
		\end{align}\label{Eq:Gamma02}
	is implemented through the substitutions
	\begin{align}
		\tilde g^{\alpha \beta} \to   G^{AB}, \qquad  \tilde \ell^{\alpha} \to i \, \mu \, \QQ^A \; , \qquad  \tilde \gamma^\alpha \to \gamma^5 \, \GA^A\, .
		\label{Eq:SubF}
	\end{align} 
	together with the set of selection rules, ($-2\epsilon$)-SRs,
		\begin{align}
		\GG^{AB}\GG^{BC} &= \GG^{AC},    & \GG^{AA}&=0,  &   \GG^{AB}&=\GG^{BA},  &
		\GA^A \GG^{AB} &= \GA^B,\nn                 \GA^A \GA^{A} &=0,  & \QQ^A \GA^{A}  &=1, &
		\QQ^A \GG^{AB} &= \QQ^B,              & \QQ^A \QQ^{A} &=1.  
		\label{Eq:2epsA}
	\end{align}
	which, ensuring the exclusion of the terms containing odd powers of $\mu$, completely
	defines the FDF and allows the construction of integrands which, upon
	integration, yield to the same result as in the FDH scheme.
	\item
	The spinors of a $d$-dimensional fermion fulfil the completeness relations 
	\begin{align}
	\sum_{\lambda=\pm}u_{\lambda}\left(\ell \right)\bar{u}_{\lambda}\left(\ell \right) & = \slashed \ell + i \mu \gamma^5 + m \, , &
	\sum_{\lambda=\pm}v_{\lambda}\left(\ell  \right)\bar{v}_{\lambda}\left(\ell \right)  & = \slashed \ell + i \mu \gamma^5  - m \, ,
	\label{Eq:CompF4}
	\end{align}
	which consistently reconstruct the numerator of the cut  propagator. 
\item In the axial gauge, the helicity sum of a $d$-dimensional transverse polarisation vector can be disentangled in
\begin{small}
	\begin{align}
	&\sum_{i=1}^{d -2} \, \varepsilon_{i\, (d)}^\mu\left (\bar \ell , \bar \eta \right )\varepsilon_{i\, (d)}^{\ast \nu}\left (\bar \ell , \bar \eta \right ) =
	\left (   - g^{\mu \nu}  +\frac{ \ell^\mu \ell^\nu}{\mu^2} \right) -\left (  \tilde g^{\mu \nu}  +
	\frac{ \tilde \ell^\mu \tilde \ell^\nu}{\mu^2} \right ) \, ,
	\label{Eq:CompGD2}
	\end{align}
\end{small}
\noindent where the first term can be regarded as the cut propagator of a massive vector boson,
\begin{align}
\sum_{\lambda=\pm,0}\varepsilon_{\lambda}^{\mu}(\ell) \, \varepsilon_{\lambda}^{*\nu}(\ell)&= -g^{\mu\nu}+\frac{\ell^{\mu}\ell^{\nu}}{\mu^{2}}  \, ,  \label{flat}
\end{align}
and the second term of the r.h.s. of Eq.~(\ref{Eq:CompGD2})  is related to the numerator of cut propagator of the scalar $s^{\bullet}$ and can be expressed in terms
of the $(-2 \epsilon)$-SRs as:
\begin{equation}
\tilde g^{\mu \nu}  +\frac{ \tilde \ell^\mu \tilde \ell^\nu}{\mu^2}  \quad \to  \quad    \GG^{AB} - \QQ^A \QQ^B  \, .
\label{Eq:Pref}
\end{equation}
\end{itemize}
Within FDF scheme, the QCD  $d$-dimensional Feynman rules in axial gauge have the following four-dimensional formulation:
\begin{subequations}
\vspace{-0.4cm}
\begin{align}
\parbox{20mm}{
\unitlength=0.20bp%
\begin{feynartspicture}(300,300)(1,1)
\FADiagram{}
\FAProp(4.,10.)(16.,10.)(0.,){/Cycles}{0}
\FALabel(5.5,8.93)[t]{\tiny $a, \alpha$}
\FALabel(14.5,8.93)[t]{\tiny $b, \beta$}
\FALabel(10.,12.5)[]{\tiny $k$}
\FAVert(4.,10.){0}
\FAVert(16.,10.){0}
\end{feynartspicture}} &= i \, \frac{ \delta^{ab}  }{k^2 -\mu^2 +i 0}\bigg(-g^{\alpha\beta}+\frac{k^{\alpha}k^{\beta}}{\mu^2}\bigg)\,,%  \quad (\mbox{gluon}), 
\label{Eq:FRglu}\\[-3.0ex]
\parbox{20mm}{\unitlength=0.20bp%
\begin{feynartspicture}(300,300)(1,1)
\FADiagram{}
\FAProp(4.,10.)(16.,10.)(0.,){/ScalarDash}{0}
\FALabel(5.5,8.93)[t]{\tiny $a, A$}
\FALabel(14.5,8.93)[t]{\tiny $b, B$}
\FALabel(10.,12.5)[]{\tiny $k$}
\FAVert(4.,10.){0}
\FAVert(16.,10.){0}
\end{feynartspicture}} &= i \,  \,\frac{\delta^{ab}}{k^2 -\mu^2+ i0} \left(G^{AB}-Q^AQ^B\right)\, , %\quad (\mbox{scalar}),  
\\[-3.0ex]
\parbox{20mm}{\unitlength=0.20bp%
\begin{feynartspicture}(300,300)(1,1)
\FADiagram{}
\FAProp(4.,10.)(16.,10.)(0.,){/Straight}{1}
\FALabel(5.5,8.93)[t]{\tiny $i$}
\FALabel(14.5,8.93)[t]{\tiny $j$}
\FALabel(10.,12.5)[]{\tiny $k$}
\FAVert(4.,10.){0}
\FAVert(16.,10.){0}
\end{feynartspicture}} &= i \, \delta^{ij} \,\frac{ \slashed k + i \mu \gamma^5 +m }{k^2 -m^2 -\mu^2+i0}  \, ,  %\quad  (\mbox{fermion}),   
\label{Eq:FRfer} 
\end{align}
\label{Eq:FR4}
\end{subequations}

\section{One-loop amplitudes}
\label{sec:2P}
In this section we present preliminary results obtained through FDF for the leading colour-ordered one-loop helicity  amplitudes $A_{5}\left(1^{+},2^{+},3^{+},4^{+},\text{H}\right)$ and $A_{6}\left(1^{+},2^{+},3^{+},4^{+},5^{+},\text{H}\right)$ in the heavy top mass limit.\\
Besides providing the analytic values of coefficients of each master integral, we give necessary ingredients to carry out this computation.

In order to apply generalised-unitarity methods within FDF, we consider as examples the one-loop  $2 \to 2,3,4$ scattering amplitudes, where external particles can be either gluons, quarks or the Higgs boson.
\smallskip

In general, due to the decomposition formulae any massless $n$-point one-loop  amplitude  can be
decomposed in terms MIs, as follows 
\begin{multline}
A_n^{\text{1-loop}} {}= \frac{1}{(4 \pi)^{2-\epsilon}}   \,   \sum_{i<j<k<l}^{n-1} \bigg [ c_{i|j|k|l;\,0}\, I_{i|j|k|l}+c_{ij|k|l;\,0}\, I_{ij|k|l} 
+c_{ij|kl;\,0}\, I_{ij|kl}\\
+ c_{i|j|k|l;\,4}\, I_{i|j|k|l}[\mu^{4}]+c_{ij|k|l;\,2}\, I_{ij|k|l}[\mu^{2}]  
+c_{ij|kl;\,2}\, I_{ij|kl}[\mu^{2}] \bigg ]  \, .
\label{Eq:Decomposition}
\end{multline}

In Eq.~(\ref{Eq:Decomposition}), we see the first line corresponds to the cut-constructible, while the second one to the rational part. Once again, we emphasise FDF gives the full contribution to the one-loop amplitude, being no need of distinguishing those two pieces.

The coefficients $c$'s entering in the decompositions~(\ref{Eq:Decomposition}) can 
be obtained by using the generalised unitarity techniques  for quadruple~\cite{Britto:2004nc,Badger:2008cm}, 
triple~\cite{Mastrolia:2006ki,Forde:2007mi,Badger:2008cm}, and double~\cite{Britto:2005ha, Britto:2006sj, Mastrolia:2009dr} cuts.  Since internal particles are massless, the single-cut 
techniques~\cite{Kilgore:2007qr,Britto:2009wz, Britto:2010um} are not needed for this computation.  In general, the cut  $C_{i_1\cdots i_k}$, defined by the conditions $D_{i_1} =\cdots = D_{i_k}=0$,  allows for the determination of the coefficients $c_{i_1\cdots i_k; \, n}$.

\subsection{The gggggH amplitude}
We show the explicit structure of the analytic contribution of the one-loop Higgs plus five gluon all-plus amplitude. For sake of simplicity, we do not write the coefficients of the finite part of the amplitude.

\noindent The leading-order contribution of the six-point can be written as,
\begin{align}
A_{6,H}^{\text{tree}}\left(1^{+},2^{+},3^{+},4^{+},5^{+},H\right)=\frac{-i\,m_{H}^{4}}{\langle1|2\rangle\langle2|3\rangle\langle3|4\rangle\langle4|5\rangle\langle5|1\rangle}.
\end{align}
\begin{figure}[htb!]
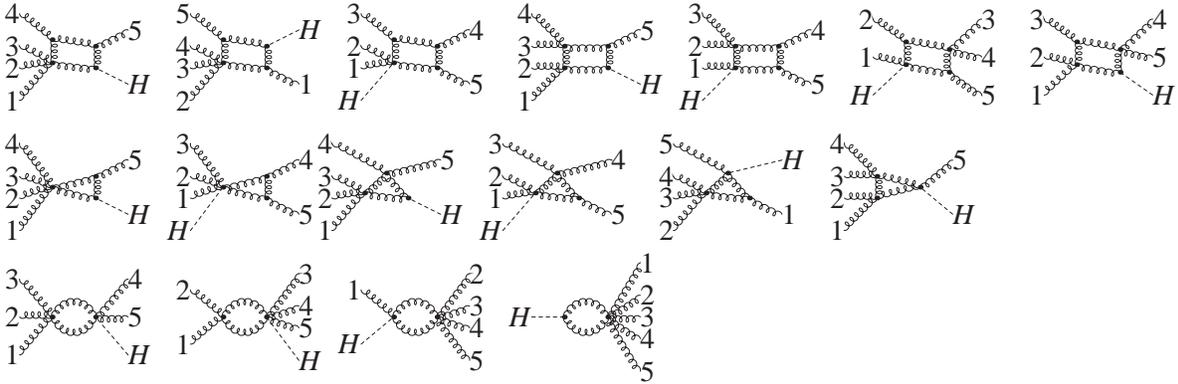

\begin{subequations}
\begin{align*}
% Boxes
&\parbox{15mm}{\input{FeynmanDiagrams/5gHQ1a.tex}}\qquad
\parbox{15mm}{\input{FeynmanDiagrams/5gHQ1b.tex}}\qquad
\parbox{15mm}{\input{FeynmanDiagrams/5gHQ1c.tex}}\qquad
\parbox{15mm}{\input{FeynmanDiagrams/5gHQ2a.tex}}\qquad
\parbox{15mm}{\input{FeynmanDiagrams/5gHQ2b.tex}}\qquad
\parbox{15mm}{\input{FeynmanDiagrams/5gHQ3a.tex}}\qquad
\parbox{15mm}{\input{FeynmanDiagrams/5gHQ3b.tex}}
\\
% Triangles
&\parbox{15mm}{\input{FeynmanDiagrams/5gHT1a.tex}}\qquad
\parbox{15mm}{\input{FeynmanDiagrams/5gHT1b.tex}}\quad
\parbox{15mm}{\input{FeynmanDiagrams/5gHT2a.tex}}\qquad
\parbox{15mm}{\input{FeynmanDiagrams/5gHT2b.tex}}\qquad
\parbox{15mm}{\input{FeynmanDiagrams/5gHT2c.tex}}\qquad
\parbox{15mm}{\input{FeynmanDiagrams/5gHT3.tex}}\\
% Bubbles
&\parbox{15mm}{\input{FeynmanDiagrams/5gHD1a.tex}}\qquad
\parbox{15mm}{\input{FeynmanDiagrams/5gHD2a.tex}}\qquad
\parbox{15mm}{\input{FeynmanDiagrams/5gHD2b.tex}}\qquad
\parbox{15mm}{\input{FeynmanDiagrams/5gHD3.tex}}
\end{align*}
\end{subequations}
\caption{Independent box-, triangle- and bubble- integral topologies for the amplitude $A_6^{1-loop}(H,1,2,3,4,5)$}\label{6top}
\end{figure}
The one-loop correction to this amplitude is obtained by considering the independent topologies depicted in fig.~\ref{6top} and takes the form, 
\begin{align}
A_{6}^{1-loop}\left(H,1^+,2^+,3^+,4^+,5^+\right)=&\frac{1}{2}A_{6}^{tree}\left(s_{1234}s_{1235}-s_{123}m_{H}^{2}\right)\,I_{123|4|5|H}\left[1\right]-\frac{1}{2}A_{6}^{tree}s_{34}s_{45}\,I_{H12|3|4|5}\left[1\right]\nn
&-\frac{1}{2}A_{6}^{tree}\left(s_{234}s_{345}-s_{34}s_{2345}\right)\,I_{H1|2|34|5}+\frac{1}{2}A_{6}^{tree}\left(s_{1234}-m_{H}^{2}\right)\,I_{1234|5|H}\left[1\right]\nn
&+c_{123|4|H|5}I_{123|4|H|5}[\mu^{4}]+c_{H12|3|4|5}I_{H12|3|4|5}[\mu^{4}]\nn
&+c_{1234|5|H}I_{1234|5|H}[\mu^{2}]+c_{1234|H|5}I_{1234|H|5}[\mu^{2}]+c_{H123|4|5}I_{H123|4|5}[\mu^{2}]\nn
&+c_{H12|34|5}I_{H12|34|5}[\mu^{2}]+c_{123|4|5H}I_{123|4|5H}[\mu^{2}]+c_{123|4H|5}I_{123|4H|5}[\mu^{2}]\nn
&+c_{12|345H}I_{12|345H}[\mu^{2}]+c_{123|45H}I_{123|45H}[\mu^{2}]+c_{H1|2345}I_{H1|2345}[\mu^{2}]\nn
&+\text{cyclic perm,}
\end{align}
being $c$'s non-vanishing coefficients.\\
A similar study was carried out for the analytic expression of the one-loop Higgs plus four gluon amplitudes , wherein the helicity configurations $A_{5}\left(H,1^{+},2^{+},3^{+},4^{+}\right)$, $A_{5}\left(H,1^{-},2^{+},3^{+},4^{+}\right)$,  $A_{5}\left(H,1^{-},2^{-},3^{+},4^{+}\right)$ and $A_{5}\left(H,1^{-},2^{+},3^{-},4^{+}\right)$ have been considered, finding agreement with~\cite{{Badger:2006us,Badger:2009hw}}\\

The procedure for computing the one-loop amplitudes given above has
been fully automated. In particular, we have implemented the FDF Feynman rules (including
the $(-2\epsilon)$-SRs) in \Feynarts/\Feyncalc\,~\cite{Hahn:2000kx}, in order to automatically build  the
tree-level amplitudes to be sewn in the cuts. Then, the coefficients of the master integrals are determined by applying integrand reduction via Laurent expansion~\cite{Mastrolia:2012bu}, which has been implemented in Mathematica, by using the package \sam~\cite{Maitre:2007jq}.
\section{Colour-Kinematics-duality}
\label{sec:3P}

In this section we briefly describe the diagrammatic study we give to the C/K-duality we presented in \cite{Mastrolia:2015maa}, where off-shell tree-level currents are embedded in higher-multiplicity/multi-loop amplitudes. This investigation is carried out by considering the tree-level diagrams for the process $gg\to X$, for massless final states, with $X=ss, q\bar{q},gg$, in four dimensions. The same analysis, first made in $d$-dimensions, has then been extended to dimensionally regulated amplitudes, taking the FDF as a regularisation scheme. The calculation has been performed in axial gauge, describing scalars in the adjoint representation and fermions in the fundamental one.
\subsection{Colour-kinematics duality for gluons}
\label{CKg}
\begin{figure}[h]
\centering
\includegraphics[scale=1.15]{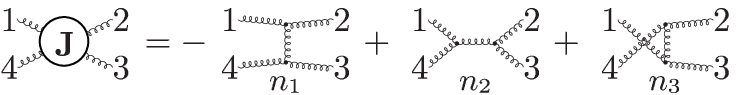}
\caption{Jacobi combination for gluons.}\label{BCJs}
\end{figure}
\par\noindent We consider the tree-level scattering, $gg\to gg$, which receives contributions from four Feynman diagrams, however, because of colour algebra, the contribution from the $4$-gluon vertex can be absorbed in the $s,t$ and $u$ channels, so that the amplitude is exposed in terms of three diagrams. The corresponding numerators, say $n_1$, $n_2$ and $n_3$, can be combined in Jacobi-like fashion as shown in Fig.~\ref{BCJs}, 
\begin{align}
N_{\text{g}}=-n_1+n_2+n_3, \label{BCJ}
\end{align}
The numerator of the gluon propagator in axial gauge has the form
\begin{equation}
\Pi^{\mu\nu}(p,q)=\Pi^{\mu\nu}_{\text{Fey}}+\Pi^{\mu\nu}_{\text{Ax}}(p,q),
\end{equation}
where $\Pi^{\mu\nu}_{\text{Fey}}$ corresponds to the numerator of the propagator in Feynman gauge and $\Pi^{\mu\nu}_{\text{Ax}}(p,q)$ labels the term depending on an arbitrary light-like reference momentum $q^{\mu}$,
\begin{align}
\Pi^{\mu\nu}_{\text{Fey}}&=-i\,g^{\mu\nu}\,, &
\Pi^{\mu\nu}_{\text{Ax}}(p,q)&=i\,\frac{p^\mu q^\nu+q^\mu p^\nu}{q\cdot p}.
\end{align}
The explicit form of (\ref{BCJ}) is given by the contraction of an
off-shell current with gluon polarisations as,
\begin{align}
\left(N_{\text{g}}\right)_{\alpha_1...\alpha_4}=
\big(J^{\mu_{1}..\mu_{4}}_{\text{g-Fey}}+J^{\mu_{1}...\mu_{4}}_{\text{g-Ax}}\big) 
\varepsilon_{\mu_1}\left(p_1,q_1\right)
\varepsilon_{\mu_2}\left(p_2,q_2\right)
\varepsilon_{\mu_3}\left(p_3,q_3\right)
\varepsilon_{\mu_4}\left(p_4,q_4\right),
\label{BCJtg}
\end{align}
where $J_{\text{s-Fey}}^{\mu_{1}\mu_{4}}$ is the sum of the Feynman gauge-like terms of the three numerators,
\begin{align}
-i\,{J}_{\text{g-Fey}}^{\mu_{1}\mu_{2}\mu_{3}\mu_{4}}(p_1,p_2,p_3,p_4)
=&p_{1}^{\mu_{1}}[g^{\mu_{3}\mu_{4}}\left(p_{1}+2p_{4}\right)^{\mu_{2}}-g^{\mu_{2}\mu_{4}}\left(p_{1}+2p_{4}\right)^{\mu_{3}}+g^{\mu_{2}\mu_{3}}\left(p_{1}+2p_{3}\right)^{\mu_{4}}]\nn
&+\text{cyclic perm.}
\label{jfeyg}
\end{align}
and
\begin{multline}
-i\, J_{\text{g-Ax}}^{\mu_{1}\mu_{2}\mu_{3}\mu_{4}}(p_{1},p_{2},p_{3},p_{4})=\frac{1}{q\cdot(p_{1}+p_{2})}\Bigg\{\\
\left(p_{1}^{\mu_{1}}p_{1}^{\mu_{2}}-p_{2}^{\mu_{2}}p_{2}^{\mu_{1}}-\left(p_{1}^{2}-p_{2}^{2}\right)g^{\mu_{1}\mu_{2}}\right)[q\cdot\left(p_{4}-p_{3}\right)g^{\mu_{3}\mu_{4}}-\left(p_{3}+2p_{4}\right)^{\mu_{3}}q^{\mu_{4}}+\left(p_{4}+2p_{3}\right)^{\mu_{4}}q^{\mu_{3}}]\\+\left(p_{3}^{\mu_{3}}p_{3}^{\mu_{4}}-p_{4}^{\mu_{3}}p_{4}^{\mu_{4}}-\left(p_{3}^{2}-p_{4}^{2}\right)g^{\mu_{3}\mu_{4}}\right)[q\cdot\left(p_{1}-p_{2}\right)g^{\mu_{1}\mu_{2}}+\left(p_{1}+2p_{2}\right)^{\mu_{1}}q^{\mu_{2}}-\left(p_{2}+2p_{1}\right)^{\mu_{2}}q^{\mu_{1}}]\Bigg\}
\\-[(1234)\to(4123)]-[(1234)\to(4231)].
\label{jaxg}
\end{multline}
is the contribution depending on the reference momentum.

\begin{figure}[htb!]
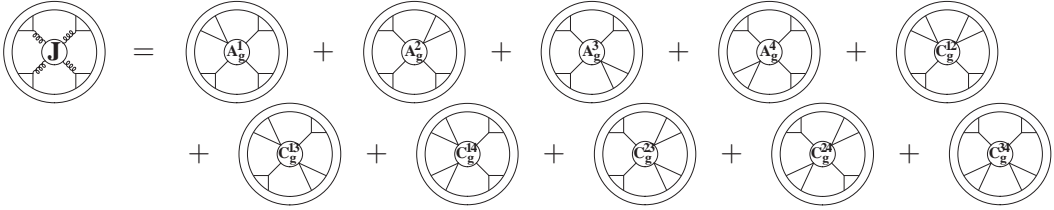

\begin{align*}
\parbox{8mm}{\input{FeynmanDiagrams2/4g.tex}}\quad\quad&=\quad
\parbox{8mm}{\input{FeynmanDiagrams2/4gP2bg.tex}}\quad\quad+\quad
\parbox{8mm}{\input{FeynmanDiagrams2/4gP3bg.tex}}\quad\quad+\quad
\parbox{8mm}{\input{FeynmanDiagrams2/4gP4bg.tex}}\quad\quad+\quad
\parbox{8mm}{\input{FeynmanDiagrams2/4gP1bg.tex}}\quad\quad+\quad
\parbox{8mm}{\input{FeynmanDiagrams2/4g12g.tex}}\\ 
\\[-2.5ex]
&\qquad+\quad
\parbox{8mm}{\input{FeynmanDiagrams2/4g13g.tex}}\quad\quad+\quad
\parbox{8mm}{\input{FeynmanDiagrams2/4g14g.tex}}\quad\quad+\quad
\parbox{8mm}{\input{FeynmanDiagrams2/4g23g.tex}}\quad\quad+\quad
\parbox{8mm}{\input{FeynmanDiagrams2/4g24g.tex}}\quad\quad+\quad
\parbox{8mm}{\input{FeynmanDiagrams2/4g34g.tex}}
\end{align*}
\caption{Off-shell colour-kinematics duality for gluons. The Jacobi combination of tree-level numerators (l.h.s)  is expressed in terms of subdiagrams only (r.h.s.). }
\label{bcjLoopa}
\end{figure}

With the expressions of the off-shell currents of  eqs.~(\ref{jfeyg},\ref{jaxg}) we embed the Jacobi-like combination of tree-level numerators into a generic diagram, where in the most general case the legs $p_1,p_2,p_3$ and $p_4$ become internal lines and polarisations associated to the particles are replaced by the numerator of their propagators.
Accordingly, Eq.~\eqref{BCJ} generalises to the following contraction,
\begin{align}
N_{\text{g}} &=(N_{\text{g}})_{\alpha_1\hdots\alpha_4}X^{\alpha_1\hdots\alpha_4}.
\label{eq:Ns:def}
\end{align}
between the tensor $(N_{\text{g}})_{\alpha_1\hdots\alpha_4}$, defined as,
\begin{equation}
(N_{\text{g}})_{\alpha_1\hdots\alpha_4}=-
\big(J^{\mu_{1}\hdots\mu_{4}}_{\text{g-Fey}}+J_{\text{g-Ax}}^{\mu_{1}\hdots\mu_{4}}\big)
\Pi_{\mu_1\alpha_1}\!\!\left(p_1,q_1\right)
\Pi_{\mu_2\alpha_2}\!\!\left(p_2,q_2\right)
\Pi_{\mu_3\alpha_3}\!\!\left(p_3,q_3\right)
\Pi_{\mu_4\alpha_4}\!\!\left(p_4,q_4\right),
\label{BCJgN}
\end{equation}
and the arbitrary tensor $X^{\alpha_1\hdots\alpha_4}$, standing for the residual kinematic dependence of the diagrams, associated to either higher-point tree-level or to multi-loop topologies. 

Using momentum conservation, we find that the r.h.s. of \eqref{BCJgN} can be cast in the following suggestive form,
\begin{align}
\left(N_{\text{g}}\right)_{\alpha_1...\alpha_4}\!&=
\sum_{i=1}^{4}p_i^2(A^i_g)_{\alpha_1...\alpha_4}+\sum_{\substack{i,j=1\\
		i\neq j}}^{4}p_i^2p_j^2(C^{ij}_g)_{\alpha_1...\alpha_4},
\label{BCJgN1}
\end{align}
where $A^{i}_g$ and $C^{ij}_g$ are tensors depending both on the momenta $p_i$ of gluons, eventually depending on the loop variables, and on the reference momenta $q_i$ of each gluon propagators.

Remarkably, Eq.(\ref{BCJgN1}) shows the full decomposition of a combinations of generic numerators built from the Jacobi relation in terms of squared momenta of the particles entering the Jacobi identity defined in Fig.~\ref{BCJs}. In particular,  this result implies that the C/K duality is certainly satisfied when imposing the on-shell cut-conditions $p_i^2 =0$. A diagrammatic representation of the consequences of the decomposition (\ref{BCJgN1}) in (\ref{eq:Ns:def}) is given in Fig.~\ref{bcjLoopa}, where each term in the r.h.s. contains an effective vertex, associated with the C/K violating term, which we have fully identified. Similar results have been obtained for $ss$ and $q\bar q$ in the final states.

\subsection{Colour-kinematics duality in $d$-dimensions}
In this Section we study the C/K-duality for tree-level amplitudes in
dimensional regularisation by employing the FDF scheme, recently
introduced in~\cite{Fazio:2014xea}.

	\begin{figure}[htb]
	\centering
		\includegraphics[scale=1.05]{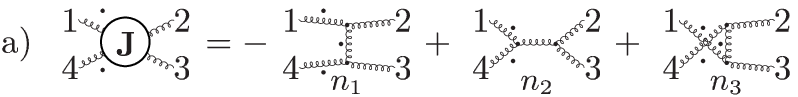}
		\includegraphics[scale=1.05]{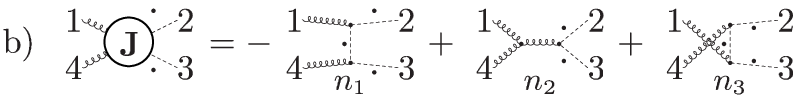}
		\includegraphics[scale=1.05]{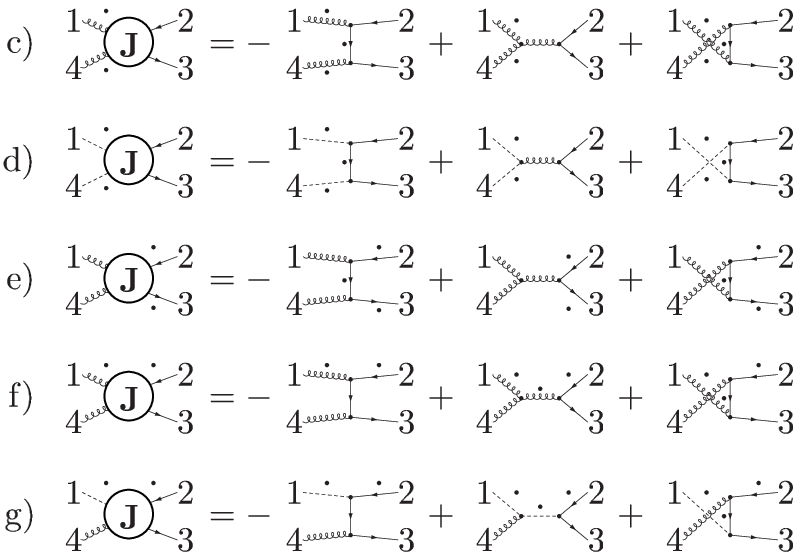}
		\caption{Jacobi combinations for FDF particles.}
		\label{BCJFDF}
	\end{figure}
The relations depicted in Fig.~\ref{BCJFDF} are the basic building blocks for the determination of higher-order scattering amplitudes within generalised unitarity based methods. The particles with dots represents the generalised (or dimensional regulated) particles, while all other lines identify particles living in $4$-dimensions. 

\noindent A diagrammatic analysis, similar to the one done in section~\ref{CKg}  was carried out in~\cite{Mastrolia:2015maa}, where the C/K duality is shown to be obeyed by the numerators of tree-level amplitudes within the FDF scheme, which involve non-trivial relations involving both massless and massive particles. The C/K-duality is recovered once transversality conditions of (generalised) gluon polarisations and Dirac equation are taken into account. More specifically, generalised gluons of momentum $p$ obey $\varepsilon_\lambda\cdot p=0\,(\lambda=0,\pm)$, while for generalised quarks one has $\bar{u}(p)(\slashed p-i\mu\gamma_5)=0$ and $(\slashed p-i\mu\gamma_5)v(p)=0$.

As a non-trivial example we provide explicit expressions for C/K dual building block for the process $g^{\bullet}g^{\bullet}(s^{\bullet}s^{\bullet})\to q\bar q g$, where initial states of gluons are treated  $d$-dimensional particles, whereas the final state remains fully four-dimensional.

\section{Conclusions}
\label{sec:4}
At one-loop level, we have made use of the unitarity methods and the Four-Dimensional-Formulation scheme to compute the analytical expressions for $gg\to ggH$ and $gg\to gggH$. 

Possible applications and extensions of this study are the computation of the full analytical expression for $\text{Higgs}+3$ jets in the final state, and as well as the two-loop implementation of the Four-Dimensional-Formulation scheme. 

On the the Colour-Kinematics side we have explicitly shown that any higher-point/loop diagram obtained from the Jacobi identity between kinematics numerators of off-shell diagrams -constructed with Feynman rules in axial gauge-
 can be decomposed in terms of sub-diagrams where one or two internal propagators are pinched. As consequence of this decomposition, we diagrammatically that proved the colour-kinematics duality is satisfied at multi-loop only by imposing on-shellness of the four particles entering in the Jacobi combination. This behaviour holds for $d$-dimensional regulated amplitudes.

\section*{Acknowledgements}
The author would like to thank Tiziano Peraro for useful discussions.
This work is supported by Fondazione Cassa di Risparmio di Padova e Rovigo (CARIPARO) and 
by Padua University Project CPDA144437.
\\The Feynman diagrams depicted in this paper were generated using \Feynarts~\cite{Hahn:2000kx}.

\bibliographystyle{utphys}
\bibliography{references}

\end{document}